\documentclass{amsart}
\usepackage{amsfonts}
\usepackage{amscd}

\input{tcilatex}
\theoremstyle{definition}
\theoremstyle{remark}
\numberwithin{equation}{section}

\input tcilatex

\begin{document}
\title{The Nondemolition Measurement of Quantum Time}
\author{V P Belavkin}
\address{Mathematics Department, University of Nottingham, NG7 2RD}
\email{vpb@maths.nott.ac.uk}
\author{M G Perkins}
\date{Received June 11,1996}
\thanks{Published in: \textit{International Journal of Theoretical Physics} 
\textbf{37} (1998) No 1, 219 - 226.}
\maketitle

\begin{abstract}
The problem of time operator in quantum mechanics is revisited. The unsharp
measurement model for quantum time based on the dynamical system-clock
interaction, is studied. Our analysis shows that the problem of the quantum
time operator with continuous spectrum cannot be separated from the
measurement problem for quantum time.
\end{abstract}

\section{Introduction: The time operator.}

The problem of time measurement in quantum theory cannot be solved within
the von Neumann theory \cite{1} simply by defining the corresponding
self-adjoint operator as a generator of the shift for the energy of a
physical system $\mathrm{S}$ (assumed to have a positive spectrum, $%
\varepsilon \in \mathbb{R}_{+}$), as no such operator exists in the Hilbert
space $\mathcal{H}_{\mathrm{S}}$.

For the purpose of simplicity let us study this problem for a quantum system
with a continuous (unbounded) energy spectrum of constant degeneracy; the
case of a free quantum particle see \cite{Hol}. The system Hilbert space can
be decomposed into a family of eigenspaces $\mathcal{H}_\varepsilon $ , $%
(\varepsilon \in \mathbb{R}_{+})$ of fixed energy. The dimensionality of $%
\mathcal{H}_\varepsilon $ corresponds to the degeneracy of the eigenvectors
corresponding to $\varepsilon $. We represent the state vectors $\psi \in 
\mathcal{H}_{\mathrm{S}}$ by a family $\left\{ \psi \left( \varepsilon
\right) \left| \varepsilon \geq 0\right. \right\} $ of Hilbert space vectors 
$\psi \left( \varepsilon \right) \in \mathcal{H}_\varepsilon $ such that $%
\int_0^\infty \left\| \psi \left( \varepsilon \right) \right\| ^2\mathrm{d}%
\varepsilon =1$.

Now, without loss of generality, we can treat all $\psi \left( \varepsilon
\right) $ as elements of some Hilbert space $\mathcal{H}$ . This is because
all the $\mathcal{H}_{\varepsilon }$ can each be embedded in the same $%
\mathcal{H}$, so for all $\varepsilon $, $\psi \left( \varepsilon \right)
\in \mathcal{H}_{\varepsilon }\subseteq \mathcal{H}$. Then we can describe
each state vector $|\psi \rangle $ by an analytic (on the upper half-plane,
where $\func{Im}(\tau )>0$) function $h:\mathbb{C\mapsto }\mathcal{H},$: 
\begin{equation*}
h(\tau )=\frac{1}{\sqrt{2\pi \hbar }}\int_{0}^{\infty }e^{\mathrm{i}%
\varepsilon \tau /\hbar }\psi \left( \varepsilon \right) \mathrm{d}%
\varepsilon .
\end{equation*}%
which is completely defined by its value on $\mathbb{R}$. This analytic
representation restricted to $\tau \in \mathbb{R}$ is called the time
representation. The Hilbert space $\widetilde{\mathcal{H}}_{S}$ of these
analytic functions with the squared norm $\left\Vert h\right\Vert
^{2}:=\int_{-\infty }^{\infty }\left\Vert h(\tau )\right\Vert ^{2}\mathrm{d}%
\tau =\left\Vert \psi \right\Vert ^{2}$ can be considered as one half of the
Hilbert space $L_{\mathcal{H}}^{2}\left( \mathbb{R}\right) $ of all
square-integrable functions of $\mathbb{R}$ with values in $\mathcal{H}$ .

In this enlarged space $L_{\mathcal{H}}^2\left( \mathbb{R}\right) $ there
exists a self-adjoint operator $\hat{\tau}$ defined by the multiplication $%
\left[ \hat{\tau}h\right] (\tau )=\tau h(\tau )$ with the eigen-spectral
family $\left\{ \mathrm{E}_t:t\in \mathbb{R}\right\} $ of orthoprojectors $%
\left[ \mathrm{E}_th\right] (\tau )=1_t(\tau )h(\tau )$ where $1_t(\tau
)=0,t\leq \tau $, and $=$ $1,t>\tau $.

However, the operator $\widehat{\tau }$ does not leave the physical subspace 
$\widetilde{\mathcal{H}}_{\mathrm{S}}\subset L_{\mathcal{H}}^{2}(\mathbb{R)}$
invariant. Instead, the unitary operator $U_{\lambda }=e^{\mathrm{i}\lambda 
\hat{\tau}/\hbar }$ functions as an isometry $h(\tau )\mapsto e^{\mathrm{i}%
\lambda \tau /\hbar }h\left( \tau \right) $ on $\widetilde{\mathcal{H}}_{S}$
corresponding to the shift $\left\vert \varepsilon \right\rangle \mapsto $ $%
\left\vert \varepsilon +\lambda \right\rangle $ on $\mathcal{H}_{\mathrm{S}%
\text{ }}$ for each $\lambda >0$. Note that the isometry on $\widetilde{%
\mathcal{H}}_{S}$ is adjoint not to $U_{\lambda }^{-1}$ but to the energy
shift operator $V_{\lambda }$ in $\widetilde{\mathcal{H}}_{\mathrm{S}},$
given by $\left[ V_{\lambda }\psi \right] \left( \varepsilon \right) =\psi
\left( \varepsilon +\lambda \right) $. The operator $V_{\lambda }^{\dagger }$
in this representation acts as 
\begin{equation*}
\left[ V_{\lambda }^{\dagger }\psi \right] \left( \varepsilon \right)
:=\left\{ 
\begin{array}{c}
\psi \left( \varepsilon -\lambda \right) ,\varepsilon >\lambda  \\ 
0,\varepsilon \leq \lambda 
\end{array}%
\right. .
\end{equation*}%
Indeed, if $\psi \left( \varepsilon \right) =\left( 2\pi \hbar \right)
^{-1/2}\int_{-\infty }^{\infty }e^{-\mathrm{i}\varepsilon \tau /\hbar
}h\left( \tau \right) \mathrm{d}\tau =0$ for all $\varepsilon <0$, then 
\begin{equation*}
\sqrt{2\pi \hbar }\left[ e^{\mathrm{i}\lambda \hat{\tau}/\hbar }h\right]
(\tau )=\int_{0}^{\infty }e^{\mathrm{i}(\varepsilon +\lambda )\tau /\hbar
}\psi \left( \varepsilon \right) \mathrm{d}\varepsilon =\int_{0}^{\infty }e^{%
\mathrm{i}\varepsilon \tau /\hbar }\left[ V_{\lambda }^{\dagger }\psi \right]
\left( \varepsilon \right) \mathrm{d}\varepsilon ,
\end{equation*}%
i.e. $e^{\mathrm{i}\lambda \widehat{\tau }/\hbar }h$ is analytic in the
upper half-plane so the operator $U_{\lambda }$ leaves $\widetilde{\mathcal{H%
}}_{S}$ invariant. Note that the shift operator $V_{\lambda }=\mathrm{P}%
_{0}U_{\lambda }^{-1}$ given by the orthoprojector $\mathrm{P}_{0}$ in $L_{%
\mathcal{H}}^{2}(\mathbb{R)}$ onto $\widetilde{\mathcal{H}}_{\mathrm{S}}=L_{%
\mathcal{H}}^{2}\left( \mathbb{R}_{+}\right) $, is defined on $\mathcal{H}_{%
\mathrm{S}\text{ }}$ by 
\begin{equation*}
V_{\lambda }\left\vert \varepsilon \right\rangle =\left\{ 
\begin{array}{c}
\left\vert \varepsilon -\lambda \right\rangle ,\lambda \leq \varepsilon  \\ 
0,\lambda >\varepsilon 
\end{array}%
\right. ,\qquad \left[ \mathrm{P}_{\lambda }\psi \right] \left( \varepsilon
\right) =\left\{ 
\begin{array}{c}
\psi \left( \varepsilon \right) ,\varepsilon >\lambda  \\ 
0,\varepsilon \leq \lambda 
\end{array}%
\right. ,
\end{equation*}%
Here $\mathrm{P}_{\lambda }=V_{\lambda }^{\dagger }V_{\lambda }$ is the
orthoprojector, giving the kernel $I-\mathrm{P}_{\lambda }$ for $V_{\lambda }
$. So the operator $V_{\lambda }$ is not isometric but only co-isometric in $%
\mathcal{H}_{\mathrm{S}\text{ }}$ as $\left[ V_{\lambda }\psi \right] \left(
\varepsilon \right) =0$ for all $\varepsilon $ if $\psi $ is localized as $%
\psi \left( \varepsilon \right) =0$ for $\varepsilon \geq \lambda $.

Although the operators $V_\lambda $ are not normal and do not commute, they
have the over-complete analytic family $\left\{ \left| s\right) \left| \func{%
Re}\left( s\right) >0\right. \right\} $ of non-orthonormal comon
eigenvectors, given by the Laplace transform $\left| s\right) =\int_0^\infty
e^{-\varepsilon s}\left| \varepsilon \right\rangle \mathrm{d}\varepsilon $
of the generalized basis $\left\{ \left| \varepsilon \right\rangle \left|
\varepsilon \in \mathbb{R}_{+}\right. \right\} .$ The proof is as follows: 
\begin{equation*}
V_\lambda \left| s\right) =\int_0^\infty e^{-\varepsilon s}V_\lambda \left|
\varepsilon \right\rangle \mathrm{d}\varepsilon =\int_\lambda ^\infty
e^{-\varepsilon s}\left| \varepsilon -\lambda \right\rangle \mathrm{d}%
\varepsilon =e^{-\lambda s}\left| s\right) .
\end{equation*}

Let us show that the vectors $\left\vert s\right) $, labelled by complex
numbers $s$ are not orthogonal, and are normalizable only if $\func{Re}(s)>0$%
. Indeed, 
\begin{equation*}
\left( s\right\vert \left. s^{\prime }\right) =\int_{0}^{\infty
}\int_{0}^{\infty }e^{-(\varepsilon ^{\prime }\bar{s}+\varepsilon s^{\prime
})}\left\langle \varepsilon ^{\prime }\right\vert \left. \varepsilon
\right\rangle \mathrm{d}\varepsilon \mathrm{d}\varepsilon ^{\prime
}=\int_{0}^{\infty }e^{-\varepsilon (\bar{s}+s^{\prime })}\mathrm{d}%
\varepsilon 
\end{equation*}%
since $\left\langle \varepsilon ^{\prime }\right\vert \left. \varepsilon
\right\rangle =\delta (\varepsilon -\varepsilon ^{\prime })$. But $%
\int_{0}^{\infty }e^{-\varepsilon (\bar{s}+s^{\prime })}\mathrm{d}%
\varepsilon =1/\left( \bar{s}+s^{\prime }\right) $, and so the vectors are
not orthogonal. If $s=s^{\prime },$ then 
\begin{equation*}
\left( s|s\right) =\frac{1}{2\func{Re}\left( s\right) }=\frac{1}{2k}<\infty
\quad \text{ if }k>0,
\end{equation*}%
where $s=k+\mathrm{i}\hbar ^{-1}\tau .$

This family is complete in $L^{2}\left( \mathbb{R}_{+}\right) $ and hence in 
$\mathcal{H}_{\mathrm{S}}$ in the sense that every state vector $\psi \in 
\mathcal{H}_{\mathrm{S}}$ can be written as an integral span 
\begin{equation*}
\psi =\frac{1}{2\pi \mathrm{i}}\int_{-\mathrm{i}\infty }^{\mathrm{i}\infty
}\left\vert s\right) \eta \left( s^{\ast }\right) \mathrm{d}s
\end{equation*}%
along any path from $-\mathrm{i}\infty $ to $\mathrm{i}\infty $ in the
domain of analyticity of the function $\eta \left( s^{\ast }\right) $, where 
$\eta \left( s\right) =\left( s\right\vert \psi $ and $s^{\ast }=-\bar{s}$.
The completeness relation, written for each component $\psi \left(
\varepsilon \right) =\left\langle \varepsilon \right\vert \psi ,$ is simply
the inversion of the Laplace $\ast $-transform

\begin{equation*}
\eta \left( s^{\ast }\right) =\int_{0}^{\infty }\left( s^{\ast }|\varepsilon
\right\rangle \psi \left( \varepsilon \right) \mathrm{d}\varepsilon ,\qquad
\left( s^{\ast }|\varepsilon \right\rangle =e^{s\varepsilon },
\end{equation*}%
since $\psi \left( \varepsilon \right) =\left( 2\pi \mathrm{i}\right)
^{-1}\int_{-\mathrm{i}\infty }^{\mathrm{i}\infty }\left\langle \varepsilon
|s\right) \eta \left( s^{\ast }\right) \mathrm{d}s,\qquad \left\langle
\varepsilon |s\right) =e^{-s\varepsilon }$. This means that the
vector-functions $\eta \left( k+\mathrm{i}\hbar ^{-1}\tau \right) =\sqrt{%
2\pi \hbar }h\left( \tau +\mathrm{i}\hbar k\right) $ define a representation
of state vectors $\psi \in \mathcal{H}_{\mathrm{S}}$ in the space of $\ast $%
-analytic functions $\eta \left( s\right) $ with the inner product 
\begin{equation*}
\left\langle \eta ^{\prime }|\eta \right\rangle =\frac{1}{\left( 2\pi
\right) ^{2}}\int_{-\mathrm{i}\infty }^{\mathrm{i}\infty }\int_{-\mathrm{i}%
\infty }^{\mathrm{i}\infty }\frac{1}{\bar{s}+s^{\prime }}\left\langle \eta
^{\prime }\left( \overline{s}\right) \left\vert \eta \left( \overline{s}%
^{\prime }\right) \right. \right\rangle \mathrm{d}\bar{s}\mathrm{d}s^{\prime
},
\end{equation*}%
given by the kernel $2\pi /\left( \overline{s}+s^{\prime }\right) $, as it
coincides with $\left\langle \psi ^{\prime }|\psi \right\rangle
=\int_{0}^{\infty }\left\langle \psi ^{\prime }\left( \varepsilon \right)
|\psi \left( \varepsilon \right) \right\rangle \mathrm{d}\varepsilon $.
However, this inner product can also be expressed as the single integral 
\begin{equation*}
\left\langle h^{\prime }|h\right\rangle =\lim_{k\downarrow 0}\frac{1}{2\pi
\hbar }\int_{-\infty }^{\infty }\left\langle \eta ^{\prime }\left( k+\mathrm{%
i}\hbar ^{-1}\tau \right) \right\vert \left. \eta \left( k+\mathrm{i}\hbar
^{-1}\tau \right) \right\rangle \mathrm{d}\tau .
\end{equation*}%
Indeed, 
\begin{equation*}
\int_{-\infty }^{\infty }\left\Vert \eta (k+\mathrm{i}\hbar ^{-1}\tau
)\right\Vert ^{2}\mathrm{d}\tau =\int_{-\infty }^{\infty }\left(
\int_{0}^{\infty }\int_{0}^{\infty }e^{-k(\varepsilon +\varepsilon ^{\prime
})+\mathrm{i}\tau (\varepsilon -\varepsilon ^{\prime })/\hbar }\left\langle
\psi \left( \varepsilon ^{\prime }\right) \right\vert \left. \psi \left(
\varepsilon \right) \right\rangle \mathrm{d}\varepsilon \mathrm{d}%
\varepsilon ^{\prime }\right) \mathrm{d}\tau .
\end{equation*}%
Now, since $\int_{-\infty }^{\infty }e^{\mathrm{i}x\tau /\hbar }\mathrm{d}%
\tau =2\pi \hbar \delta \left( x\right) $, we obtain 
\begin{equation*}
\int_{-\infty }^{\infty }\left\Vert \eta \left( k+\mathrm{i}\hbar ^{-1}\tau
\right) \right\Vert ^{2}\mathrm{d}\tau =2\pi \hbar \int_{0}^{\infty
}e^{-2k\varepsilon }\left\Vert \psi \left( \varepsilon \right) \right\Vert
^{2}\mathrm{d}\varepsilon .
\end{equation*}%
This, given that the family of vectors $|s)$ is non-orthogonal, means that
it is over-complete. The equality is true for all $\psi $ and since $\psi
\left( \varepsilon \right) =\left\langle \varepsilon \right\vert \psi $ and $%
\eta \left( k+\mathrm{i}\hbar ^{-1}\tau \right) =\left( k+\mathrm{i}\hbar
^{-1}\tau \right\vert \psi $ then this can be written equivalently as 
\begin{equation*}
\int_{-\infty }^{\infty }\left\vert k+\mathrm{i}\hbar ^{-1}\tau \right)
\left( k+\mathrm{i}\hbar ^{-1}\tau \right\vert \mathrm{d}\tau =2\pi \hbar
\int_{0}^{\infty }e^{-2k\varepsilon }\left\vert \varepsilon \right\rangle
\left\langle \varepsilon \right\vert \mathrm{d}\varepsilon =2\pi \hbar
e^{-2kH},
\end{equation*}%
where $H$ is the induced Hamiltonian of the system in $L^{2}\left( \mathbb{R}%
_{+}\right) $ . In the limit as $k\rightarrow 0,$ we obtain 
\begin{equation*}
2\pi \hbar \left\Vert h\right\Vert ^{2}=2\pi \hbar \int_{-\infty }^{\infty
}\left\Vert h\left( \tau \right) \right\Vert ^{2}\mathrm{d}\tau =2\pi \hbar
\int_{0}^{\infty }\left\langle \psi \left( \varepsilon \right) \right\vert
\left. \psi \left( \varepsilon \right) \right\rangle \mathrm{d}\varepsilon
=2\pi \hbar \left\Vert \psi \right\Vert ^{2},
\end{equation*}%
that is $\int_{-\infty }^{\infty }\left\vert \mathrm{i}\hbar ^{-1}\tau
\right) \left( \mathrm{i}\hbar ^{-1}\tau \right\vert \mathrm{d}\tau =2\pi
\hbar \widehat{1}$.

\section{The Ideal Unsharp Measurement of Time.}

Let us consider the non-orthonormal family of right eigenvectors $\left\{
\left\vert s\right) :\func{Re}\left( s\right) =0\right\} $ for the co-shift
operators $V_{\lambda }$ at the limit $\func{Re}\left( s\right) \rightarrow 0
$. Now 
\begin{equation*}
\frac{1}{2\pi \hbar }\int_{-\infty }^{\infty }\left\vert \mathrm{i}\hbar
^{-1}\tau \right) \left( \mathrm{i}\hbar ^{-1}\tau \right\vert \mathrm{d}%
\tau =\frac{1}{2\pi \hbar }\lim_{k\downarrow 0}\int_{-\infty }^{\infty
}\left\vert k+\mathrm{i}\hbar ^{-1}\tau \right) \left( k+\mathrm{i}\hbar
^{-1}\tau \right\vert \mathrm{d}\tau =\hat{1},
\end{equation*}%
and we have the normalization condition $\int \left\Vert h\left( \tau
\right) \right\Vert ^{2}\mathrm{d}\tau =1$ if $\left\Vert \psi \right\Vert
^{2}=1$. So we can treat 
\begin{equation*}
h\left( \tau \right) =\frac{1}{\sqrt{2\pi \hbar }}\left( \mathrm{i}\hbar
^{-1}\tau \right\vert \psi =\frac{1}{\sqrt{2\pi \hbar }}\eta \left( \mathrm{i%
}\hbar ^{-1}\tau \right) 
\end{equation*}%
as the probability amplitude of a time measurement ($\tau $-measurement)
described by the continuous over-complete family of generalized vectors 
\begin{equation*}
\chi \left( \tau \right) =\frac{1}{\sqrt{2\pi \hbar }}\left\vert \mathrm{i}%
\hbar ^{-1}\tau \right) ,\qquad \tau \in \mathbb{R}.
\end{equation*}

For each Borel subset, $\triangle $, the integral $\int_{\triangle
}\left\vert \mathrm{i}\hbar ^{-1}\tau \right) \left( \mathrm{i}\hbar
^{-1}\tau \right\vert \mathrm{d}\tau $ defines the unsharp, positive,
contractive operator $\Pi _{\triangle }$ acting as 
\begin{equation*}
\Pi _{\triangle }\psi =\int_{\triangle }\chi \left( \tau \right) h(\tau )%
\mathrm{d}\tau 
\end{equation*}%
with $h\left( \tau \right) =\chi \left( \tau \right) ^{\dagger }\psi $. The
map $\Delta \mapsto \Pi _{\Delta }$ defines a positive operator-valued
measure, normalized to the identity operator: $\Pi _{\mathbb{R}_{+}}=\hat{1}$
in $\widetilde{\mathcal{H}}_{\mathrm{S}}$ and so in $\mathcal{H}_{\mathrm{S}}
$. However, this measure is not orthogonal (projector-valued) and this is
why it describes the unsharp (fuzzy) measurement of the time in initial
state $\psi $. This measurement gives the best results among all the unsharp
measurements of the time parameter of a coherent quantum signal under the
maximal likelihood criterion \cite{5,6}. However, the non-orthogonal vectors 
$\chi \left( \tau \right) $ cannot be regarded as the time eigenstates that
are not yet normalized, as they are not normalizable. Moreover, such ideal
measurements demolish the quantum system because there is no way to obtain
the \textit{a posteriori} state vectors $\psi _{\tau }\in \mathcal{H}_{%
\mathrm{S}}$, compatible with this measurement, using the projection or any
other reduction postulate. To show this, consider the generalization \cite{4}
of the projection postulate to the continuous spectrum case. This states
that after the measurement returning a result $\tau $, the state of the
system is given by the normalization $\psi _{\tau }=G\left( \tau \right)
\psi /\left\Vert G\left( \tau \right) \psi \right\Vert $ of a linear
transform $G\left( \tau \right) \psi $ of the \textit{a priori} state vector 
$\psi $ given by a family $\left\{ G\left( \tau \right) \right\} $ of
Hilbert space operators $G\left( \tau \right) $ with $\int\nolimits_{-\infty
}^{\infty }G\left( \tau \right) ^{\dagger }G\left( \tau \right) \mathrm{d}%
\tau =\hat{1}$. The operator-valued measure $\Pi _{\Delta }$ is then defined
by the integration $\Pi _{\Delta }=\int_{\Delta }G^{\dagger }\left( \tau
\right) G\left( \tau \right) \mathrm{d}\tau $ of the operator-valued density 
$\Pi \left( \tau \right) =G^{\dagger }\left( \tau \right) G\left( \tau
\right) $ of this measure. However, because of the continuity of time, $\tau 
$, there are no eigen-projectors corresponding to the continuous values, $%
\tau \in \mathbb{R}$. Therefore, instead of the orthoprojectors, some other,
non-orthogonal, reduction operators $G\left( \tau \right) $ corresponding to
an unsharp time measurement must be used to obtain a Hilbert space state
vector $G\left( \tau \right) \psi $ with $\left\Vert G\left( \tau \right)
\psi \right\Vert <\infty $ for (almost) each result $\tau $ of the
measurement.

For a candidate measurement operator $G\left( \tau \right) $ to make
physical sense, it must satisfy certain conditions. Two of these have
already been dealt with, but there still remain:

\qquad (i) The family $\left\{ G\left( \tau \right) \right\} $ must commute
with the energy coshift, $\left[ G\left( \tau \right) ,V\left( \lambda
\right) \right] =0,\lambda \in \mathbb{R}_{+}$, so that the non-demolition
measurement of time will be compatible with the ideal time measurement,
described by the time vectors $|\mathrm{i}\hbar ^{-1}\tau ),$i.e.; 
\begin{equation*}
\left( s\right\vert G\left( \tau \right) =g_{s}\left( \tau \right) \left(
s\right\vert \text{ ,\qquad }\tau \in \mathbb{R}\text{ ,}
\end{equation*}%
where $g_{s}\left( \tau \right) $ are complex $L^{2}\left( \mathbb{R}\right) 
$-functions.

\qquad (ii) The family $\left\{ G\left( \tau \right) \right\} $ must be
covariant with respect to the time shift, 
\begin{equation*}
e^{-\mathrm{i}Ht/\hbar }G\left( \tau -t\right) =e^{\mathrm{i}\theta \left(
t\right) }G\left( \tau \right) e^{-\mathrm{i}Ht/\hbar },\quad t\in \mathbb{R}%
,
\end{equation*}%
where $\theta \left( t\right) \in \lbrack 0,2\pi )$, so that the predicted
physics is unchanged by our choice of the origin for time.

One can easily show that the ideal unsharp measurement examined above is not
compatible with these conditions, because there is no such covariant $%
G\left( \tau \right) $ that commutes with $V_{\lambda }^{\dagger }$, for
which 
\begin{equation*}
\int_{\Delta }G\left( \tau \right) ^{\dagger }G\left( \tau \right) \mathrm{d}%
\tau =\Pi _{\Delta }=\frac{1}{2\pi \hbar }\int_{\Delta }\left\vert \mathrm{i}%
\hbar ^{-1}\tau \right) \left( \mathrm{i}\hbar ^{-1}\tau \right\vert \mathrm{%
d}\tau 
\end{equation*}%
for any measurable $\Delta \subset \mathbb{R},$

Suppose that this were so. We know that the probability density of $\tau $
is given by $\left\vert h\left( \tau \right) \right\vert ^{2}$. On the other
hand, from the commutivity with $G\left( \tau \right) $, it follows that the 
\textit{a posteriori} state vector $\psi _{\tau }$ is obtained by modulation
by some filter (or envelope) function $g_{s}\left( \tau \right) $ in the $s$%
-representation, and then by the normalization: 
\begin{equation*}
\eta _{\tau }\left( s\right) =\left( s\right\vert \psi _{\tau }=\frac{%
g_{s}\left( \tau \right) \eta \left( s\right) }{c\left( \tau \right) }
\end{equation*}%
where 
\begin{equation*}
\left\vert c\left( \tau \right) \right\vert ^{2}=\frac{1}{2\pi \mathrm{i}}%
\int_{\func{Re}\left( s\right) =0}\left\vert g_{s}\left( \tau \right)
\right\vert ^{2}\left\Vert \eta \left( s\right) \right\Vert ^{2}\mathrm{d}s
\end{equation*}%
is the corresponding probability density. As these two expressions for the
probability density must be equal, and since $\left\Vert \eta \left(
s\right) \right\Vert ^{2}=2\pi \hbar \left\Vert h\left( \tau \right)
\right\Vert ^{2}$, where $s=\mathrm{i}\hbar ^{-1}\tau ^{\prime }$, then $%
\left\vert g_{s}\left( \tau \right) \right\vert ^{2}$must be a delta
function. There is, however, no such square integrable function $g_{s}\left(
\tau \right) .$

\section{A Realisation of Unsharp Measurement of Time.}

The covariant measurement operators $G\left( \tau \right) $ can be obtained
from the interaction model with a clock, generalizing the model \cite{4,3}
with the discrete spectrum. The non-Hermitian model for such interaction is
similar to the model for non-demolition measurements of quantum phase \cite%
{2}. It is given by the non-unitary interaction operator $V_{-\widehat{x}}=%
\mathrm{P}_{0}e^{\mathrm{i}\left( \widehat{\tau }\otimes \widehat{x}\right)
/\hbar }$, defining $G\left( \tau \right) =\mathrm{P}_{0}\varphi \left( \tau
-\widehat{\tau }\right) $ as 
\begin{equation*}
G\left( \tau \right) \psi \left( \varepsilon \right) =\left( \langle
\varepsilon |\otimes \langle \tau |\right) V_{-\widehat{x}}\left( \psi
\otimes \varphi \right) =\psi \left( \varepsilon -\widehat{x}\right) \varphi
\left( \tau \right) .
\end{equation*}%
Here $\widehat{x}$ is the momentum operator of the clock pointer, $|\tau
\rangle ,$ $\tau \in \mathbb{R}$ are the generalized eigenvectors of the
self-adjoint operator $P:f\mapsto \mathrm{i}\hbar f^{\prime }$ in $%
L^{2}\left( \mathbb{R}\right) $, describing the continuous pointer position
in the momentum representation $\widehat{x}f\left( x\right) =xf\left(
x\right) $, and $\varphi \left( \tau \right) =\widetilde{f}\left( \tau
\right) $ is a clock wavefunction, given as the involute transform 
\begin{equation*}
\widetilde{f}\left( \tau \right) =\left( 2\pi \hbar \right)
^{-1/2}\int_{-\infty }^{\infty }\overline{f}\left( x\right) e^{\mathrm{i}%
\tau x/\hbar }\mathrm{d}x
\end{equation*}%
of the admissible initial state $f\left( x\right) =0$, $x>0$, $\left\Vert
\varphi \right\Vert ^{2}=\int_{-\infty }^{0}\left\vert f\left( x\right)
\right\vert ^{2}\mathrm{d}x=1$ (with negative pointer momentum.) Because the
admissible wavefunctions cannot be localized in the position representation,
the time measurement is always unsharp, but it can be made almost sharp by
choosing $f\left( x\right) =1/\sqrt{E}$ for $x\in \left[ -E,0\right] $ and $%
f\left( x\right) =0$ for $x\notin \left[ -E,0\right] $ and going to the
limit $E\rightarrow \infty $.

Consider as an example the cases where the continuous energy spectrum is in
the range (a) $\left[ 0,\infty \right) $ and (b) $\left[ 0,E\right] $ . Then;

(a) Take the Hilbert space of the clock to be given by $\left\{ \left|
x\right\rangle \left| x\in \left( -\infty ,0\right] \right. \right\} $ as in
the discrete case. Then consider a wavefunction of the clock in the momentum
representation of the form $f\left( x\right) =\left( 2\lambda \right)
^{\frac 12}e^{\lambda x}$ , with $\lambda >0$ real. This is normalised and
we can find the wavefunction of the pointer in the position representation,
given by $\varphi \left( \tau \right) =\left\langle \tau \right| \varphi $ ,
which is;

\begin{equation*}
\varphi \left( \tau \right) =\left( \frac{\hbar \lambda }{\pi }\right) ^{%
\frac{1}{2}}\frac{1}{\hbar \lambda +\mathrm{i}\tau }
\end{equation*}%
Hence the probability distribution of $y$ is; $\left\vert \varphi \left(
\tau \right) \right\vert ^{2}=\hbar \lambda /\pi \left( \tau ^{2}+\hbar
^{2}\lambda ^{2}\right) $.

Now since $\langle \tau |V_{-\widehat{x}}\varphi =\mathrm{P}_0\varphi \left(
\tau -\widehat{\tau }\right) $ and $\Pi \left( \tau \right) =G\left( \tau
\right) ^{\dagger }G\left( \tau \right) $ is the operator valued density for
time measure, then $\left| \varphi \left( \tau \right) \right| ^2$ gives the
probability of measuring a time different from the mean time by $y.$ There
are thus two possibilities for the sharp measurement of time:

(i) In the classical limit, $\hbar =0$ , we obtain $\left| \varphi \left(
\tau \right) \right| ^2$ as a delta function corresponding to the exact
classical measurement of time.

(ii) In the limit as $\lambda \rightarrow 0,$ we again find that $\left|
\varphi \left( \tau \right) \right| ^2$ takes the form of a delta function
and hence get sharp measurement of time. We cannot, however, take $\lambda
=0 $ but since the height of the function $\left| \varphi \left( \tau
\right) \right| ^2$ at $\tau =0$ is $\frac 1{\hbar \lambda }$ then we
effectively obtain sharp measurement of time when $\lambda \leq \frac 1\hbar
.$

(b) The clock Hilbert space is now $\left\{ \left| x\right\rangle \left|
x\in \left[ -E,0\right] \right. \right\} .$ Hence the normalised clock
wavefunction in the momentum representation becomes;

\begin{equation*}
f\left( x\right) =\left( \frac{2\lambda }{1-e^{-2\lambda E}}\right) ^{\frac
12}e^{\lambda x}
\end{equation*}
So in the position representation we obtain;

\begin{equation*}
\left| \varphi \left( \tau \right) \right| ^2=\frac{\hbar \lambda }{\pi
\left( 1-e^{-2\lambda E}\right) \left( \tau ^2+\hbar ^2\lambda ^2\right) }%
\left( 1-2e^{-\lambda E}\cos \frac{\tau E}\hbar +e^{-2\lambda E}\right)
\end{equation*}

As before, if we consider the classical limit, $\hbar =0,$ then we obtain
the sharp measurement of time. Now consider the limit $\lambda \rightarrow
0. $ In this case, we find that, when $\tau \neq 0,$

\begin{equation*}
\left| \varphi \left( \tau \right) \right| ^2\rightarrow \frac{\hbar \left(
1-\cos \frac{\tau E}\hbar \right) }{\pi E\tau ^2}
\end{equation*}
and when $\tau =0$, $\left| \varphi \left( \tau \right) \right|
^2\rightarrow \frac E{2\pi \hbar }$. Hence it is not sufficient to have $%
\lambda =0$ (which is equivalent to the clock wavefunction $f\left( x\right)
=1/\sqrt{E}$ for $x\in \left[ -E,0\right] $ and $f\left( x\right) =0$ for $%
x\notin \left[ -E,0\right] $) but we must also take $E\rightarrow \infty $
as noted above in order to obtain a sharp measurement.


\begin{thebibliography}{9}
\bibitem{1} von Neumann, J. Mathematical Foundation of Quantum Mechanics.
Princeton University Press, 1955.

\bibitem{Hol} Holevo, A. S. Probabilistic and Statistical Aspects of Quantum
Theory. Amsterdam: North Holand, 1982.

\bibitem{4} Belavkin, V.P. Nondemolition Principle of Quantum Measurement
Theory. Foundations of Physics, \textbf{24}, p 685, 1994.

\bibitem{3} Stratonovich, R.L. and Belavkin, V.P. On the Dynamical
Interpretation of the Quantum Measurement Postulate. Int. J. of Theor.
Phys., \textbf{35}(11), pp2215--2228, 1996.

\bibitem{2} Belavkin, V.P. and Bendjaballah, C. Continuous Measurements of
Quantum Phase. Quantum Opt. \textbf{6}, pp 169-186, 1994.

\bibitem{5} Helstrom C.W. Quantum Detection and Estimation Theory. Academic,
New York, 1976.

\bibitem{6} Holevo. A. S. Rep. Math. Phys. \textbf{13}(3), pp287--307, 1978.
\end{thebibliography}
\end{document}